\documentclass[reprint,amsmath,amssymb,aps]{revtex4-1}

\usepackage{color}
\usepackage{amsmath}
\usepackage{graphicx}

\begin{document}

\preprint{APS/123-QED}
\title{Parity-Induced Thermalization Gap in Disordered Ring Lattices}
\author{Yao Wang,$^{1,2,3}$ Jun Gao,$^{1,2,3}$  Xiao-Ling Pang,$^{1,3}$ Zhi-Qiang Jiao,$^{1,3}$ Hao Tang,$^{1,3,4}$ Yuan Chen,$^{1,2,3}$ Lu-Feng Qiao,$^{1,3}$ Zhen-Wei Gao,$^{1,3}$ Jian-Peng Dou,$^{1,3}$, Ai-Lin Yang$^{1,3}$}
\author{Xian-Min Jin$^{1,3,4,}$}
\email{xianmin.jin@sjtu.edu.cn}
\affiliation{
	$^1$School of Physics and Astronomy, Shanghai Jiao Tong University, Shanghai 200240, China\\
	$^2$Institute for Quantum Science and Engineering and Department of Physics, Southern University of Science and Technology, Shenzhen 518055, China\\
	$^3$Synergetic Innovation Center of Quantum Information and Quantum Physics, University of Science and Technology of China, Hefei, Anhui 230026, China\\
	$^4$Institute of Natural Sciences, Shanghai Jiao Tong University, Shanghai 200240, China}

\date{Published 8 January 2019}

\begin{abstract}
The gaps separating two different states widely exist in various physical systems: from the electrons in periodic lattices to the analogs in photonic, phononic, plasmonic systems, and even quasicrystals. Recently, a thermalization gap, an inaccessible range of photon statistics, was proposed for light in disordered structures [Nat. Phys. 11, 930 (2015)], which is intrinsically induced by the disorder-immune chiral symmetry and can be reflected by the photon statistics. The lattice topology was further identified as a decisive role in determining the photon statistics when the chiral symmetry is satisfied. Being very distinct from one-dimensional lattices, the photon statistics in ring lattices are dictated by its parity, i.e, odd or even sited. Here, we for the first time experimentally observe a parity-induced thermalization gap in strongly disordered ring photonic structures. In a limited scale, though the light tends to be localized, we are still able to find clear evidence of the parity-dependent disorder-immune chiral symmetry and the resulting thermalization gap by measuring photon statistics, while strong disorder-induced Anderson localization overwhelms such a phenomenon in larger-scale structures. Our results shed new light on the relation among symmetry, disorder, and localization, and may inspire new resources and artificial devices for information processing and quantum control on a photonic chip.
\end{abstract}

\maketitle

Symmetry is crucial for establishing energy gaps for electrons in lattices and analogs~\cite{gap_electron,gap_photonic,gap_phononic,gap_plasmonic,gap_quasicrystal}. Disorder often decreases such effects~\cite{gap_electron}, but not for all cases. Certain disorder-immune chiral symmetries~\cite{chiral_1,chiral_2,chiral_3,chiral_4} would emerge in random matrix theory~\cite{random_matrix_theory}, such as chiral~\cite{chiral_3} and particle-hole symmetric ensembles~\cite{random_matrix_theory_particle}, and play a key role in fields ranging from superconductivity~\cite{random_matrix_theory_supericonductivity} to quantum chromodynamics~\cite{random_matrix_theory_chromodynamics}. A distinctive characteristic of disorder-immune chiral symmetry is that the system Hamiltonian can be transformed into a block off-diagonal matrix form, a separate bipartite sublattices~\cite{hallmark_chiral_matrix,hallmark_chiral,random_matrix_theory}, and is disorder-immune behaving as holding for each realization from a disordered ensemble~\cite{np}. 

\begin{figure}[t!]
	\centering
	\includegraphics[width=0.88\columnwidth]{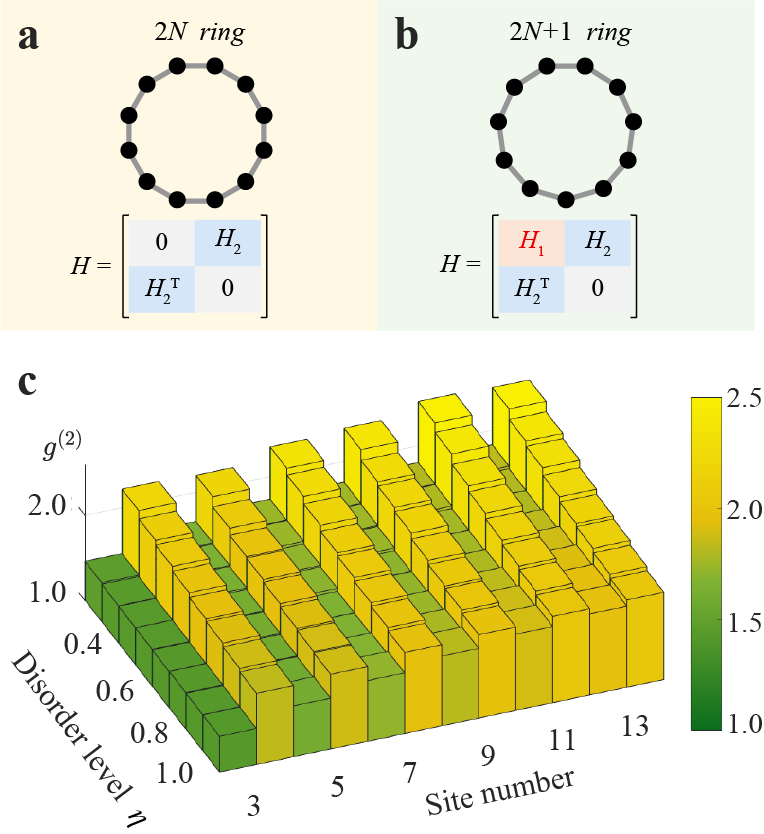}
	\caption{{\bf Hamiltonian form and intensity correlation of disordered ring lattices.} (a),(b) The Hamiltonian of off-diagonal disordered ring lattices with different site numbers. The Hamiltonian for even-sited ring lattices could be rearranged into block off-diagonal matrix form {\bf(a)}, but not for odd-sited {\bf(b)}. (c) There exists a significant gap of intensity correlation $g^{(2)}$ in even-parity ring lattices. The difference of $g^{(2)}$ between odd- and even-sited ring lattices becomes smaller and even disappears when the disorder level and the site number increase. The simulation parameters are adopted as $\bar{c} = 0.5$ and the normalized propagation distance $z\bar{c} = 17.5$.}
	\label{1}
\end{figure}

The tight-binding lattices have been suggested to study disorder-immune chiral symmetry~\cite{np}. The resulting thermalization gap, not by parity, is circuitously confirmed in one-dimensional disordered lattices~\cite{kon2016sub,kon2017}. Inspired by the works on photonic thermalization gap~\cite{np} and the effect of lattice topology~\cite{kon2017lattice}, we investigate the scenario of parity-dependent disorder-immune chiral symmetry in ring lattices, chiral symmetry for even-sited ring lattices, and chiral-symmetry breaking for odd-sited ones. The resulting thermalization gap can be revealed by observing the photon statistics in even-parity disordered ring lattices.

In this Letter, we demonstrate experimental observation of the parity-induced thermalization (PIT) gap on a photonic chip. By using femtosecond laser direct writing, we are able to freely prototype waveguides as sites of lattices and introduce disorder by precise coupling control in three dimensions. We measure the evolved light distribution out of 120 lattices with randomly picked coupling parameters in a fixed disorder level to obtain photon statistics. We observe a thermalization gap in an even-parity disordered ring lattice where the cross-correlation cannot go below a certain limit due to disorder-immune symmetries. As a way to demonstrate such parity-induced thermalization gap, we observe that the cross-correlation in even-sited lattices is significantly larger than the one in odd-sited lattices. This work may deepen the understanding of the relation among symmetry, disorder, and localization, and inspire applications for quantum integrated photonics.

The dynamics of light in photonic lattices can be described by a set of coupled discrete Schr\"{o}dinger equations~\cite{eq1_0}, which are derived from the Schr\"{o}dinger-type paraxial wave equation~\cite{eq1_1,eq1_2} by employing the tight-binding approximation~\cite{chr2003}
\begin{equation}
    -i \frac{\partial\psi_{n}}{\partial z}= c_{n-1}\psi_{n-1}+c_{n+1}\psi_{n+1}
\end{equation}
where $\psi_{n}$ is the complex field amplitude of site $n$, $z$ is the propagation distance along the waveguides and maps the time variable, and coefficient $c$ represents the coupling strength between the neighboring waveguides. The equations could also be rewritten in the matrix form: $-i\frac{\partial\psi}{\partial z}=H\psi$, where the Hamiltonian $H$ is the coupling coefficient matrix.

We consider a photonic lattice model in which the sites are arranged in a closed $ring$ structure with nearest-neighbor-only coupling. The off-diagonal disorder \cite{off_diagonal_disorder}, especially, is introduced into the lattice via randomly varying coupling coefficient $c$. Here, we consider that $c$ follows the uniform distribution between $\bar{c}-\Delta c$ and $\bar{c}+\Delta c$, where $\bar{c}$ is the average value of the coupling coefficient, $\Delta c$ is half-width of $c$ range, and defines the disorder level as $\eta=\Delta c/\bar{c}$.

\begin{figure}
	\centering
	\includegraphics[width=0.88\columnwidth]{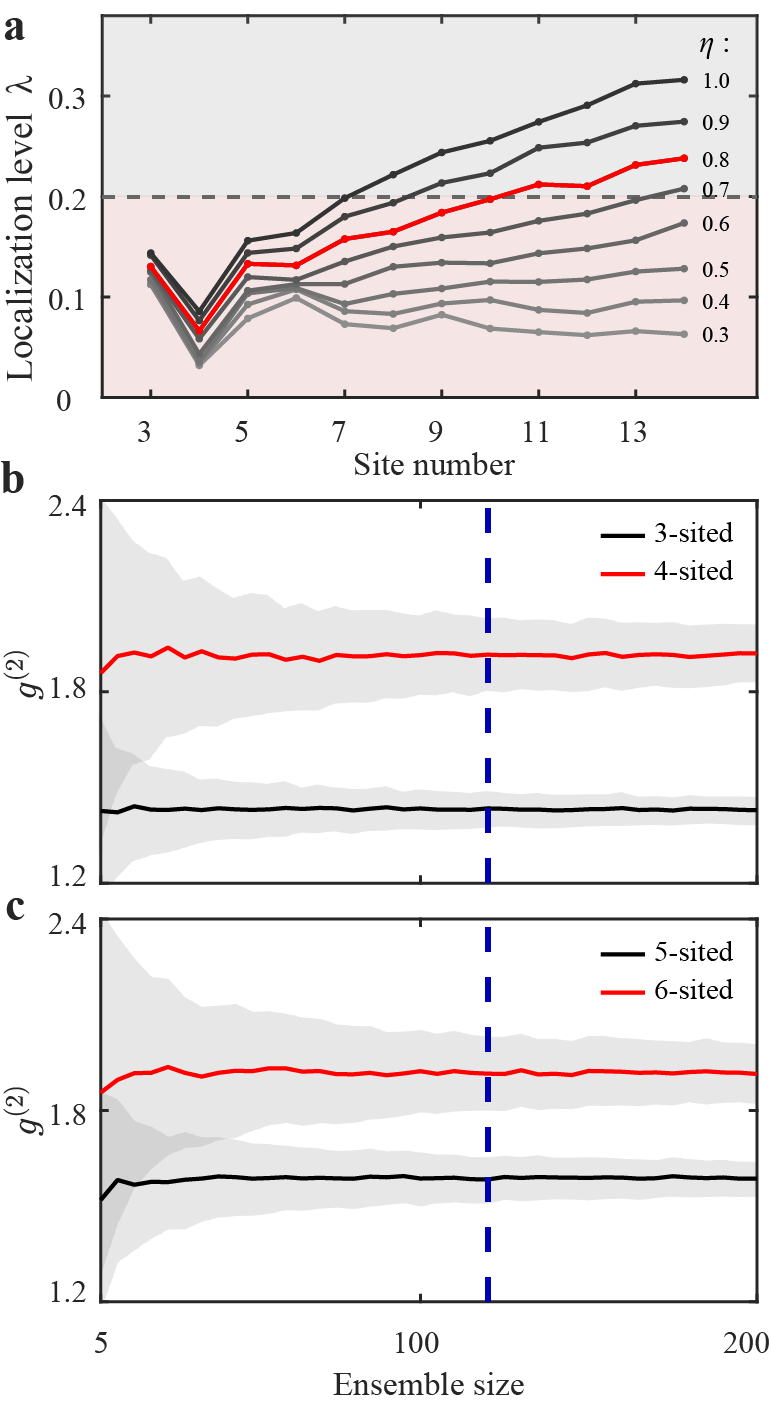}
	\caption{{\bf Theoretical simulation of the PIT gap in disordered ring lattices.} (a) The localization level varies with the disorder level and system scale. The derived bound (gray dashed line) divides the region where the thermalization gap disappears or not. The disorder level $\eta =0.8$ shown as the red line is chosen in our experiment to ensure the simultaneous observation of the thermalization gap and Anderson localization. (b),(c) The distinguishability of the thermalization gap depends on the size of the ensemble. Statistically, the gray regions showing 1 standard deviation away from the averaged $g^{(2)}$ of the odd- and even-sited ring lattices (obtained from 800 simulations) overlap with each other when the ensemble size is small, and the zones clearly separate when the sample number goes larger than 100. The red and black line is the mean value of $g^{(2)}$, and the blue-dashed line presents the ensemble size adopted in our experiment.}
	\label{2}
\end{figure}

In this model, the Hamiltonian $H$ for an even-sited ring lattice could be rearranged into a block off-diagonal matrix form [Fig.\ref{1}(a)], while the $H$ for odd-sited ring lattice cannot [Fig.\ref{1}(b)]. It means that the disorder-immune chiral symmetry emerges in even-sited ring lattices, but not in odd-sites ring lattices.
The interplay between disorder and symmetry could be revealed through observing the high-order statistics: the normalized intensity correlation of light after evolution, corresponding to the photon statistics $g^{(2)}$ in quantum optics theory, would be larger when the disorder-immune chair symmetry is satisfied. The $g^{(2)}$ can be obtained by measuring the light intensity $I$~\cite{np}
\begin{equation}
    g^{(2)}=\frac{\langle I^2\rangle}{\langle I\rangle^2}
\end{equation}
where $\langle \cdot \rangle$ denotes ensemble average over disorder realizations. Apparently, the $g^{(2)}$ is uniformly equal to $1$ for all nondisordered system.

\begin{figure}
	\centering
	\includegraphics[width=0.98\columnwidth]{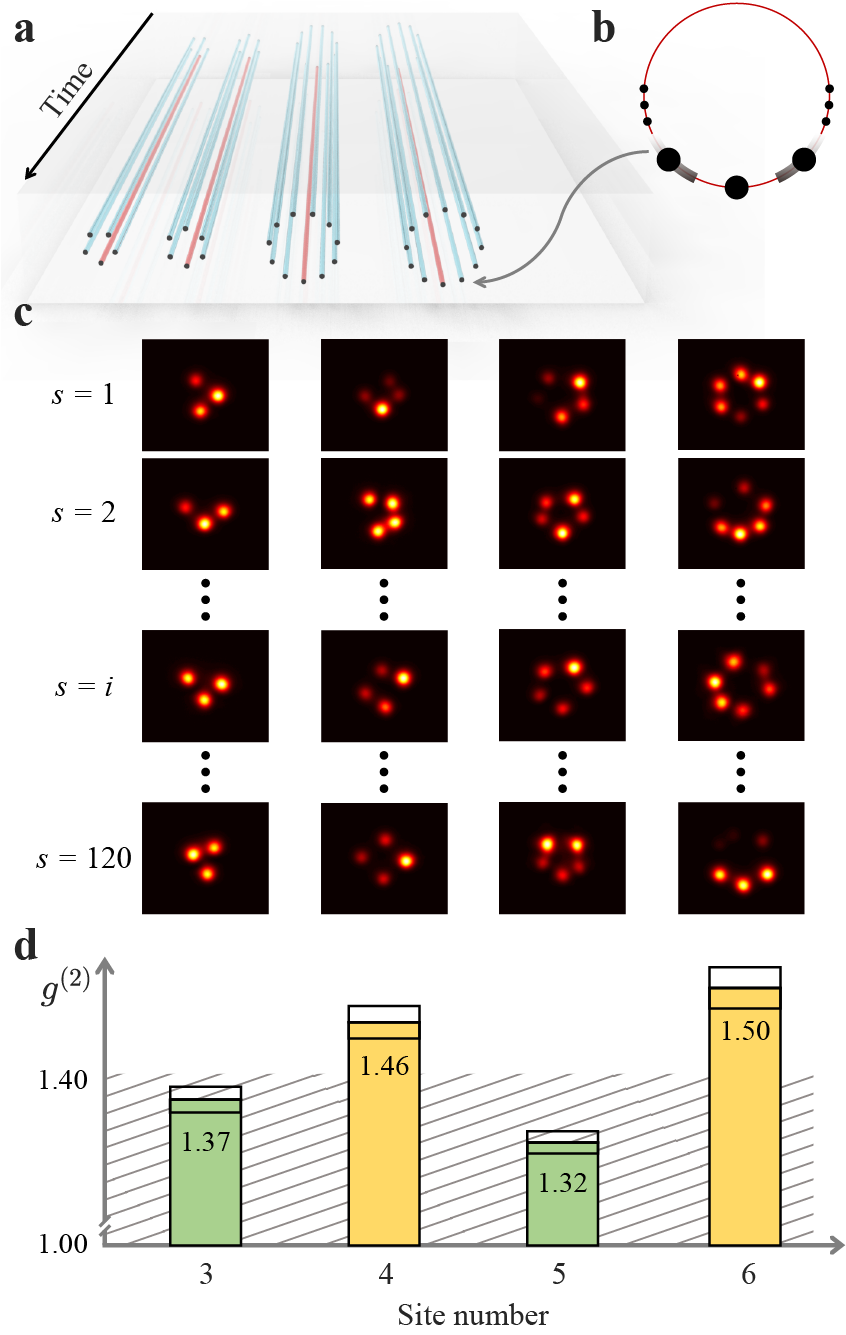}
	\caption{{\bf Sketch of ring lattices and experimental observation of the PIT gap.} (a) The ring lattice chips are fabricated by using the femtosecond laser direct writing technique. The propagation distance $z$ along waveguides in the ring lattices maps the time variable. (b) Sketch of introducing uniform-probability-distributed disorder. The shade of gray denotes the probability of position distribution, which is not uniform due to the exponential relation between the separation distance and coupling strength. (c) Experimentally imaging the output intensity distributions for samples up to 120. (d) The $g^{(2)}$ of even-sited ring lattices is distinctly larger than the value of odd-sited ring lattices. The dashed area presents the thermalization gap ranging from 1.0 to 1.4.} 
	\label{3}
\end{figure}

We investigate the dynamics under this lattice model, and the result shows there is indeed a thermalization gap in even-parity ring lattices. In Fig.\ref{1}(c), the simulation result of $g^{(2)}$ for the excited sites is illustrated. At a low disorder level, the $g^{(2)}$ of even-sited ring lattices is substantially larger than the one of odd-sited ring lattices, representing an existence of thermalization difference between them. The thermalization difference tends to be smaller and even disappear when the disorder level and the system scale increase, see the lower right area in Fig.\ref{1}(c). The physical mechanism behind is that the strong Anderson localization behavior prevents the light from spreading and therefore nullifies the impact of lattice size and its parity~\cite{np,kon2017lattice}.

\begin{figure*}
	\centering
	\includegraphics[width=1.88\columnwidth]{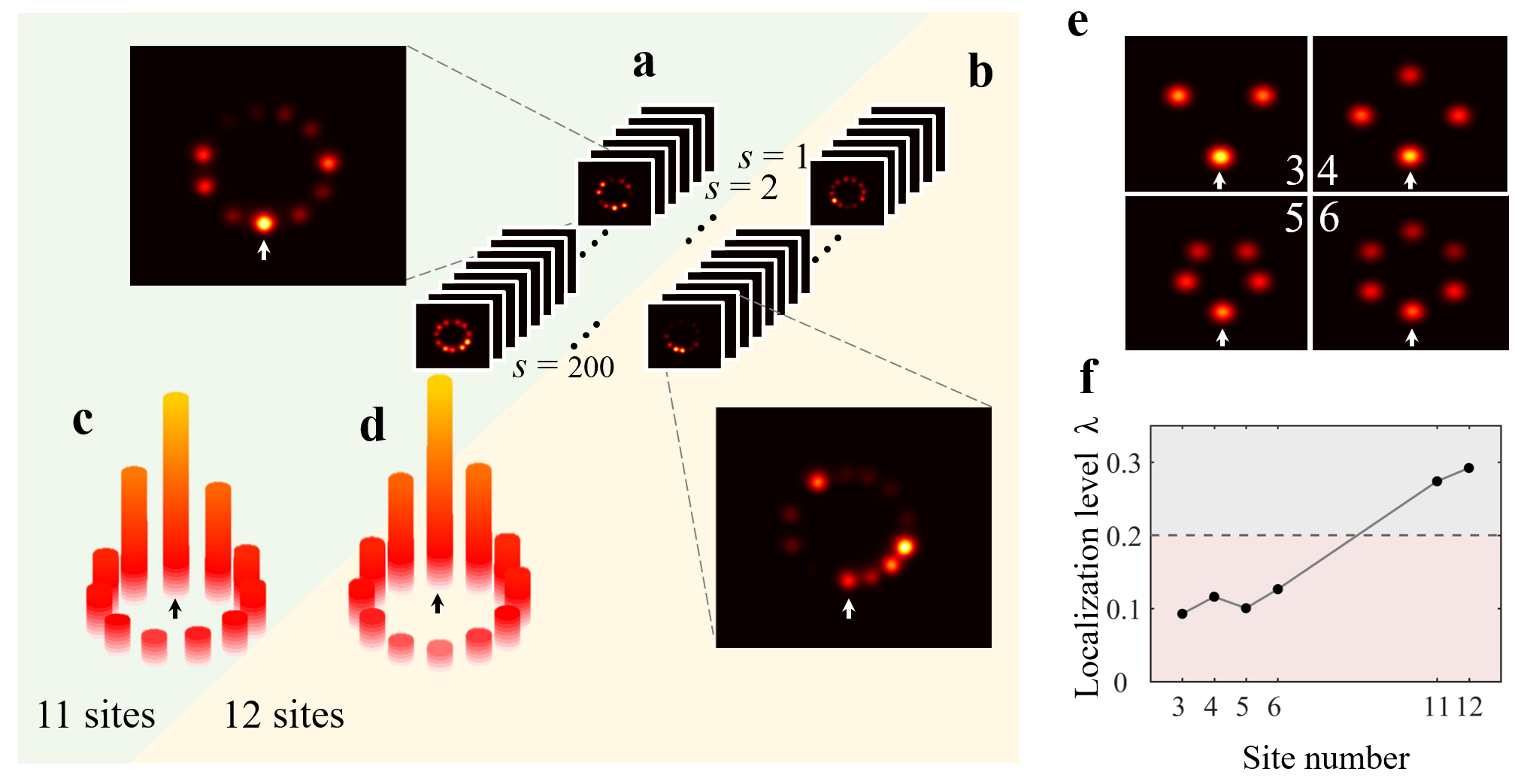}
	\caption{{\bf Experimental observation of Anderson localization.} (a),(b) The imaged output intensity distributions of 11- and 12-sited ring lattices. For each sample, the light tends to be localized in one or two sites. (c),(d) The averaged results of 11- and 12-sited ring lattices show that the light is more likely to be localized in the excited site and nearest-neighboring sites. (e) The averaged intensity distributions of 3-, 4-, 5- and 6-sited ring lattices. The light tends to be equally localized to all the sites. (f) The measured localization level agrees well with theoretical prediction. The black/white arrows mark the excited sites.}
	\label{4}
\end{figure*}

The relation among system scale, disorder, and localization level can be well visualized in Fig.\ref{2}(a). We define $\lambda = \frac{1}{2}\sum _1^N|I_i-1/N|$ to quantify the localization level, where $N$ is the site number. We identify a bound of 0.2 to divide the regions for being able to observe the thermalization gap and localization respectively (see Supplemental Material~\cite{SM1}). We can see that the disorder must be large enough to enable the observation of the two phenomena while the ultimate large value is not accessible in practice. Our simulation suggests that a disorder level of 0.8 is preferable since we can observe the thermalization gap for the site number under 6 and localization for the site number above 11.

Besides, the size of the ensemble seriously impacts the distinguishability of the gap. As a statistical result, the thermalization difference between odd- and even-sited ring lattices appears only when the size of the ensemble is large enough. As shown in Fig.\ref{2}(b), \ref{2}(c), the $g^{(2)}$ uncertainty range of odd- and even-sited ring lattices overlaps when the ensemble size is small, and significantly reduces as the ensemble size increases. As long as the sample number becomes larger than 100, the expected values of $g^{(2)}$ tend to be stable and their uncertainty ranges are substantially suppressed so that the thermalization difference becomes clearly distinguishable. The results imply that more than hundreds of ring lattices have to be fabricated and measured for every fixed-site number to confirm the disorder-immune chiral symmetry emerging; this is very experimentally challenging, not only due to the huge workload but also repeatability of fabrication and testing systematic parameters.

In our experiment, we construct the ring lattices in a borosilicate glass wafer using the femtosecond direct laser writing technique [Fig.\ref{3}(a)] \cite{fabriol,fab1,fab2,GJ,fab3}. We introduce $\eta =0.8$ off-diagonal disorder into the ring lattices, with average coupling $\bar{c}\approx 0.5$ mm$^{-1}$ at the light wavelength of $\lambda = 852$ nm and the normalized propagation distance $z\bar{c}\approx 17.25$. The uniform-probability-distributed random coupling coefficients $c$ between the two sites are realized by precisely controlling the relative positions of the waveguides (see Supplemental Material~\cite{SM2}). As shown in Fig.\ref{3}(b), the position distribution of each site is not uniform due to the exponential relation between the separation distance and coupling strength \cite{TH}. We achieve the aforementioned demanding requirement on repeatability by monitoring and locking the parameters of the laser, the chip carrier, and the lab environment with a closed loop. We fabricate 120 and 200 ring lattices for each site number to observe the thermalization gap and the strong Anderson localization behavior, respectively. 

In Fig.\ref{3}(c), we display part of the experimentally imaged output intensity distribution. We can see that the light tends to be localized but still spreads to all the sites. We can derive the intensity correlation $g^{(2)}$ with all these experimental data (see Supplemental Material~\cite{SM3}). The obtained correlations of even-sited ring lattices in which the disorder-immune chiral symmetry is satisfied, $1.46\pm 0.019$ for 4 sited and $1.50\pm 0.024$ for 6 sited, are distinctly larger than the values of odd-sited ring lattices, $1.37\pm 0.015$ for 3 sited and $1.32\pm 0.013$ for 5 sited (see Fig.\ref{3}(d)). We can see that there is a distinct difference of thermal behavior in photon statistics between the odd- and even-sited ring lattices. The $g^{(2)}$ of the even-parity ring lattice cannot go below 1.40, which implies that there is a clear thermalization gap ranging from $g^{(2)} = 1.0$ to 1.4 in even-parity ring lattices. The disorder-immune chiral symmetry and induced thermal gap are not directly visible from the individual output density image, but can be revealed by probing the high-order statistics. Surprisingly, albeit that the imperfection of fabrication and measurement may happen over 480 samples, the predicted photonic thermalization gap still can be successfully observed, revealing the robustness of the disorder-immune chiral symmetry. 

As has been shown in Fig.\ref{2}(a), with the same disorder level of 0.8, we will be able to observe that Anderson localization emerges while the thermalization gap disappears for the site number above 11. The intensity distributions of 11- and 12-sited ring lattices are shown in Fig.\ref{4}(a),\ref{4}(b). The light behaves in a strongly localized manner, and it is worth mentioning that the specifically localized position of the light can be whichever site, not necessarily at the excited site. After averaging all sample results, the strong disorder-induced Anderson localization is even more distinct. In a finite evolution time, the light is more likely to be localized in the excited site and nearest-neighboring sites [Fig.\ref{4}(c),\ref{4}(d)] for example, the proportions of the excited site are up to 0.24 and 0.25 for 11- and 12-sited ring lattices, respectively. 

Figure.\ref{4}(e) shows the averaged intensity distributions of 3-, 4-, 5-, and 6-sited ring lattices. In contrast, the light spreads to all the sites and the proportion of every site is nearly uniform though the light tends to be localized in every individual sample. We calculate the localization level $\lambda$ for the experimental data, and the result agrees well with the simulation, that is, $\lambda < 0.2$ for 3-, 4-, 5- and 6-sited ring lattices and $\lambda > 0.2$ for 11- and 12-sited ring lattices [see Fig.\ref{4}(f)]. Interestingly, with the same disorder level, Anderson localization is weak in limited-scale lattice structures, but becomes strong and breaks the disorder-immune chiral symmetry in large scale.  

In summary, we experimentally observe the PIT gap in strongly disordered ring photonic lattices. We identified the existence of disorder-immune chiral symmetry in even-sited disordered ring lattices by measuring photon statistics, manifesting an intensity correlation dependence on the parity of site number. 

Our results have interpreted a delicate relation among symmetry, disorder, and localization. The PIT gap is clear in a limited-scale structure, while the disorder-induced Anderson localization becomes dominant in a larger-scale structure. Though the symmetry is normally sensitive to the disorder, it can become more robust when chiral symmetry is satisfied. Chiral symmetry can still be broken by a large disorder, especially in large-scale structures. 

With the successful experimental implementation here, many fascinating questions related to entropy generation can be explored in the future \cite{np}, for example, the potential impact of nonlinearities induced in the lattice, the dynamics with a varied time dimension, and the evolution of nonclassical light in activated chiral lattices. It is also intriguing to investigate the existence of disorder-immune symmetry and the associated thermalization gap in quasicrystals \cite{gap_quasicrystal,QC} and the Hubbard model \cite{AA1,AA2,AA3}.

\subsection*{Acknowledgments}
The authors thank Jian-Wei Pan for helpful discussions. This work was supported by National Key R\&D Program of China (2017YFA0303700); National Natural Science Foundation of China (NSFC) (61734005, 11761141014, 11690033); Science and Technology Commission of Shanghai Municipality (STCSM) (15QA1402200, 16JC1400405, 17JC1400403); Shanghai Municipal Education Commission (SMEC)(16SG09, 2017-01-07-00-02-E00049); X.-M.J. acknowledges support from the National Young 1000 Talents Plan.\\

\section*{Supplemental Material}
\subsection*{\large I. The bound of observing thermalization gap and Anderson localization}
\begin{figure*}[ht!]
	\centering
	\includegraphics[width=1.6\columnwidth]{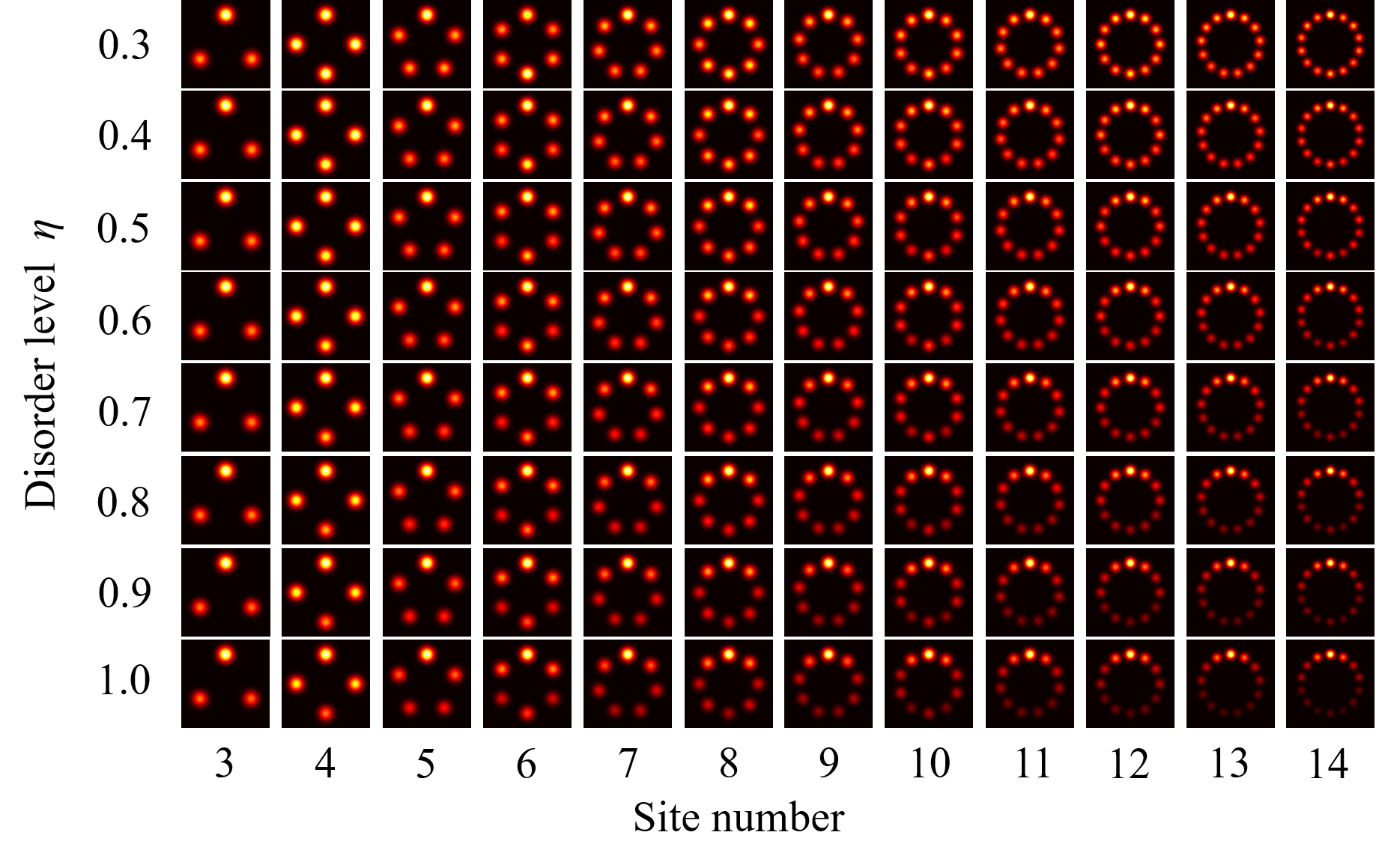}
	\caption{{\bf Averaged distribution of light varying with the disorder level and site number.} The photons in disordered ring lattice tend to be localized with the site number and the disorder level increasing.}
	\label{s1}
\end{figure*}

\begin{figure*}[ht!]
	\centering
	\includegraphics[width=1.2\columnwidth]{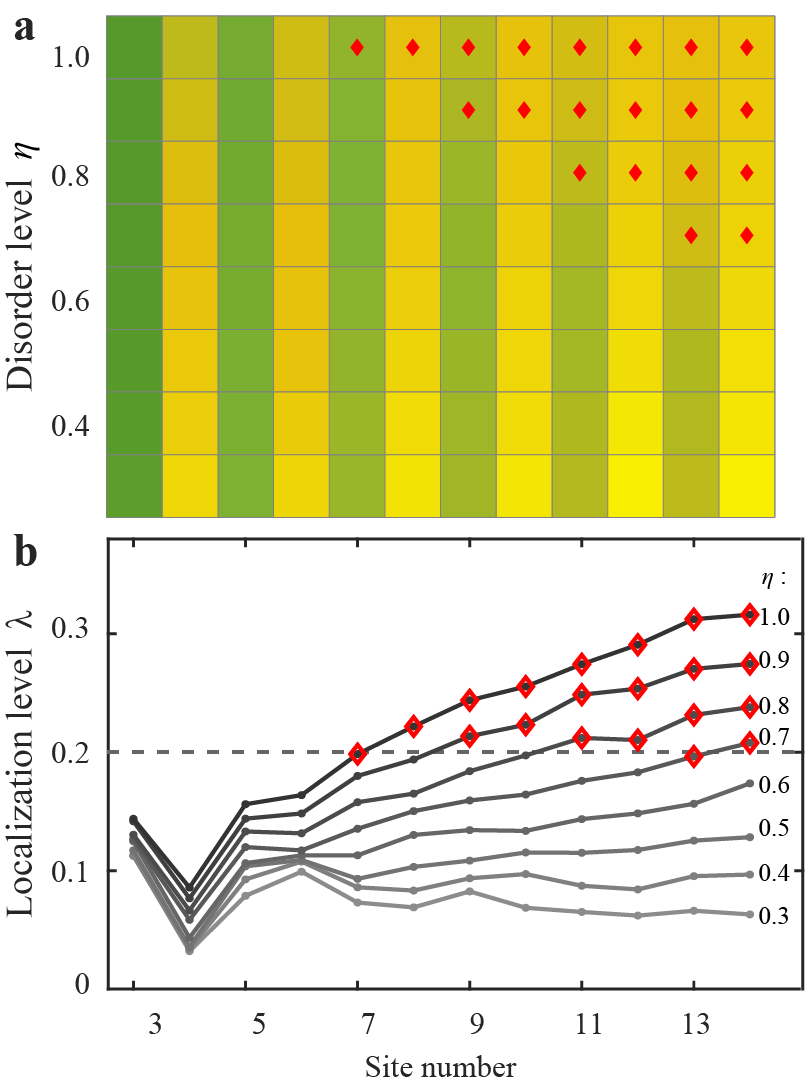}
	\caption{{\bf The derivation of the bound of observing thermalization gap and Anderson localization.} {\bf a.} The red rhombus mark the area where the difference of $g^{(2)}$ between odd- and even-sited ring lattices is smaller than 0.3. {\bf b.} We mark the corresponding position which give a bound around 0.2.}
	\label{s2}
\end{figure*}

It should be noticed that the localization level defined in previous work [S1] can not be directly applied to this work. Their definition of localization level is not sensitive to the site number. With each fixed site number, the averaged distributions of light and their trend as a function of disorder level can be visually seen in Fig.\ref{s1} corresponding to the results presented in Fig.2(a) in main text. To identify the bound of dividing the regions for being able to observe thermalization gap and localization respectively, we present the results of the simulated $g^{(2)}$ and the scale of localization level together. As shown in Fig.\ref{s2}(a), we mark the area with the red rhombus where the difference of $g^{(2)}$ between odd- and even-sited ring lattices is smaller than 0.3, which is hard to be distinguished in theory due to the statics fluctuation. We then mark the corresponding position in Fig.\ref{s2}(b) which give a bound around 0.2.

\subsection*{\large II. Introducing uniform-probability-distributed disorder} 
Fig.3(b) in main text illustrates how to introduce uniform-probability-distributed disorder. We determine each site's position before performing fabrication. Firstly we determine the separation distance between each two nearest-neighboring sites according to characterized coupling coefficient, then calculate and fix the circumscribed circle since all sites supposed to stay on a circle, then we will be able to fix the position of every site. The average separation distance is 9.76$\ \mu m$ corresponding to $\bar{c}\approx 0.5\ mm^{-1}$ at the light wavelength of $\lambda = 852\ nm$. The position of each site is specifically determined so that the coupling strength is uniformly distributed with the disorder level $\eta=0.8$. It should be noticed that the relation between the separation distance and the coupling strength is not linear. In every lattice, the waveguides are designed to be 34.5 $mm$ long, corresponding to the normalized propagation distance $z\bar{c}\approx 17.25$, to ensure that the light has hoped sufficiently before being measured.

\subsection*{\large III. Fabrication and measurement of on-chip ring lattices} 
We design the lattice structure according to the characterized coupling coefficients modulated by the separation between two adjacent waveguides. The lattices are written in borosilicate glass (refractive index $n_0=1.514$) with femtosecond laser (10W, 1026nm, 290fs pulse duration, 1MHz repetition rate and 513nm working wavelength). Before the laser writing beam is focused inside the borosilicate substrate with a 50X objective lens (numerical aperture of 0.55), we control the shape and size of the focal volume of the beam with a cylindrical telescope. A high-precision three-axis motion stage is used to move the photonic chip during fabrication with a constant velocity of 5mm/s.

\begin{figure*}[ht!]
	\centering
	\includegraphics[width=1.6\columnwidth]{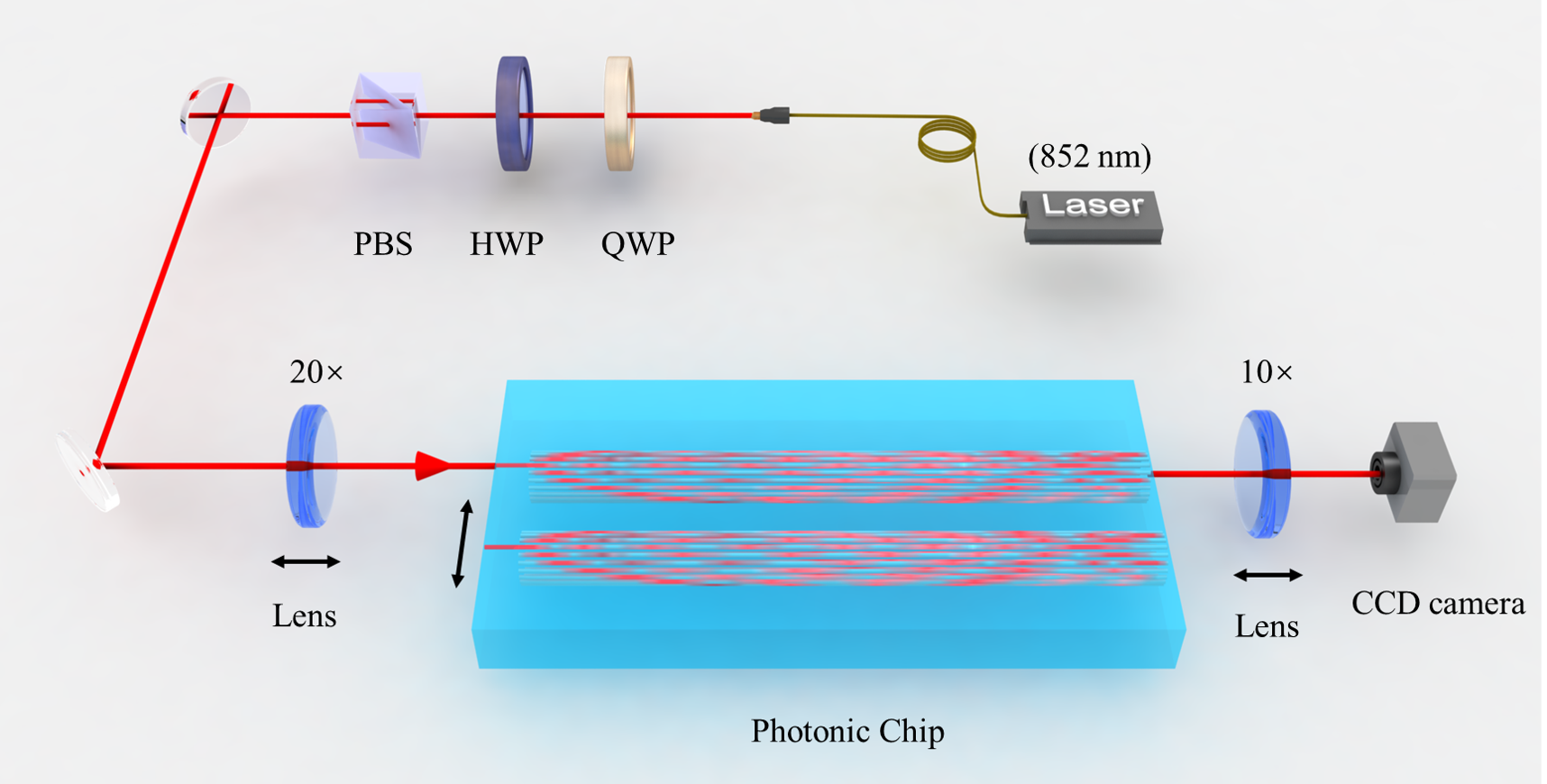}
	\caption{{\bf Experimental setup.} The photons with a wavelength of $852\ nm$ and a power of $1\ mW$ are prepared in horizontal polarization via a polarizing beam splitter, and then are injected into the Entry waveguides. A quarter-wave plate and half-wave plate are employed to compensate the polarization rotations of fiber. Each photonic chip contains 40 samples. The black arrows note the directions of lens and photonic chip could be tuned.}
	\label{s3}
\end{figure*}

The schematic of experimental setup is shown in Fig.\ref{s3}. Experiments are performed by focusing the horizontal polarization coherent photons into the Entry waveguide in the photonic chip using a 20X objective lens. The evolution output from the lattices is observed using a 10X microscope objective lens and the CCD camera after propagation within the lattices. 

\subsection*{\large IV. Obtaining intensity distribution from CCD camera}
The intensity distribution of output photons is recorded by CCD camera with format of image and data. Part of images are shown in Fig.3 \& 4 in the main text. We analyze the data to obtain the light intensity of each site with the position information which is determined before performing fabrication. The spot size is determined with same ratio of edge ($1/e$) to center value under the Gaussian fitting. Since all the unwanted output pattern of the scattered light is much weaker than the edge intensity, such that the intensity distribution can be obtained free of background noises. 

\subsection*{\large V. Deriving the intensity correlation $g^{(2)}$}
To faithfully retrieve the intensity correlation $g^{(2)}$, the condition of statistical stationarity needs to be satisfied: the total density at the output facet of every sample is uniform, which means it is necessary to normalize the output density before calculating $g^{(2)}$. The next challenge is to give a preciser statistical uncertainty from 120 samples. We process the experiment data with Monte Carlo method. We randomly select $N$ samples from 120 experiment results, then calculate the $g^{(2)}_1$ of the $N$ samples selected according to the equation (2). We repeat the steps above for $M$ times. After those operations, there are $M$ results: $g^{(2)}_1, g^{(2)}_2\dots g^{(2)}_i\dots g^{(2)}_M$. We set the average value $\bar{g}^{(2)}=\frac{1}{M}\sum _i g^{(2)}_i $ as the final experiment $g^{(2)}$ result and the standard deviation is the uncertainty.

\subsection*{\large VI. Experimental amplitude distributions over ensemble average}
\begin{figure*}[ht!]
	\centering
	\includegraphics[width=2\columnwidth]{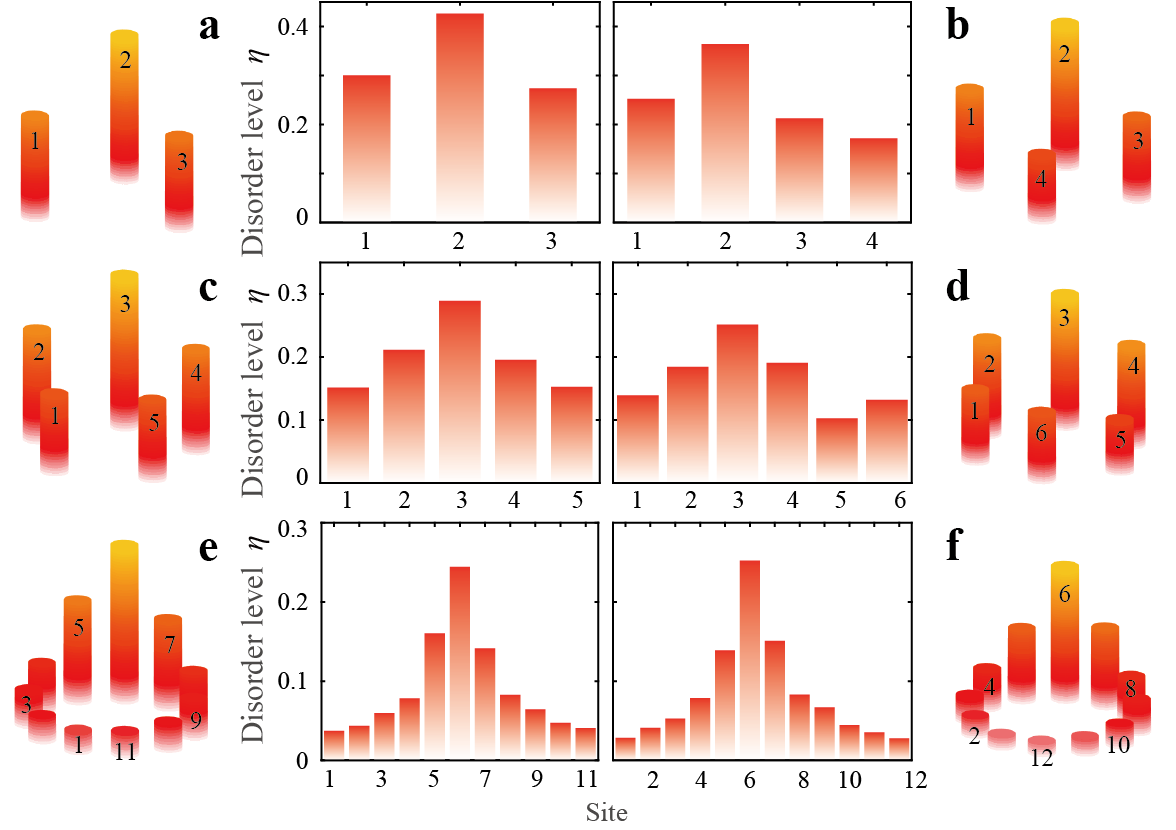}
	\caption{{\bf Experimental amplitude distributions over ensemble average.} The distributions of 11/12-sited lattices ({\bf e, f}) localized obviously higher than that of 3- to 6-sited lattices  ({\bf a-d}), which can also be truly reflected by relevant index we defined. The site label 2, 3 and 6 are the excited sites for 3-(4-), 5-(6-) and 11-(12-)sited ring lattice respectively.}
	\label{distribution}
\end{figure*}

In main text, we show the ensemble-average amplitude distributions of 11/12-sited and 3- to 6-sited ring lattices with different forms in Fig.4(c-e). The reason behind is that the form in Fig.4(c-d) is unsuitable for small-sited ring lattice with fine visibility (The histogram figures for 3- to 6-sited ring lattices can not well eye-guiding the existence of rings). However, we still illustrate all scenarios together in the same form here. To compare the relative localization fairly and confirm visually the relevant index for localization level we defined, we re-exhibit the ensemble-average amplitude distributions with same form in Fig.\ref{distribution}. The distributions of 11/12-sited lattices localized obviously higher than that of 3- to 6-sited lattices, which can also be truly reflected by relevant index we defined.

\subsection*{\large VII. Experimental amplitude distribution and the role of Anderson localization} 
\begin{figure*}[ht!]
	\centering
	\includegraphics[width=2\columnwidth]{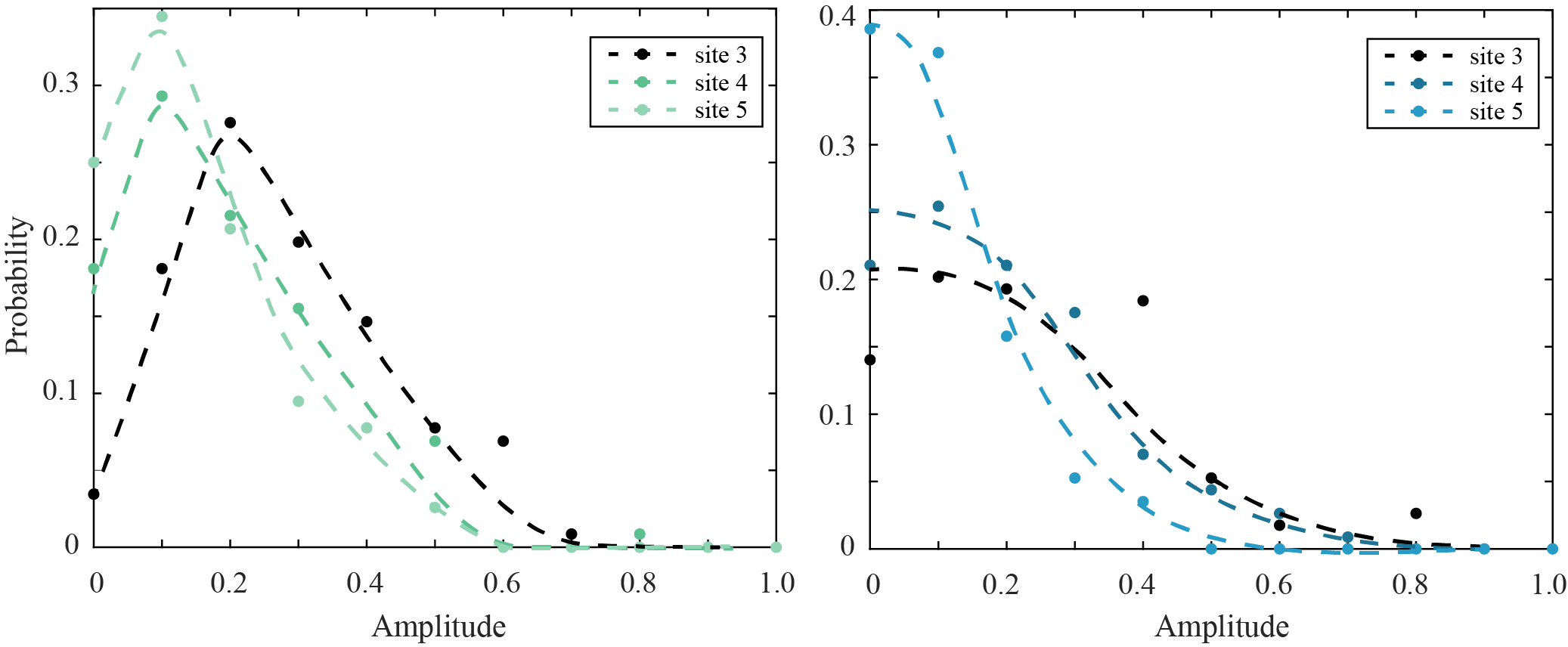}
	\caption{{\bf Probability distributions of amplitude at different output sites for 5-sited (left) and 6-sited (right) lattices.} The odd-sited ring lattices have a Rayleigh-like amplitude distribution for sites in the lattice due to the non-chiral symmetry, while the even-sited lattices exhibit Gaussian amplitude distributions. The site 3 is the excited site, and the sites 4 and 5 are the neighboring and the next neighboring sites respectively.}
	\label{s_5_6_distribution_each_site}
\end{figure*}

\begin{figure*}[ht!]
	\centering
	\includegraphics[width=2\columnwidth]{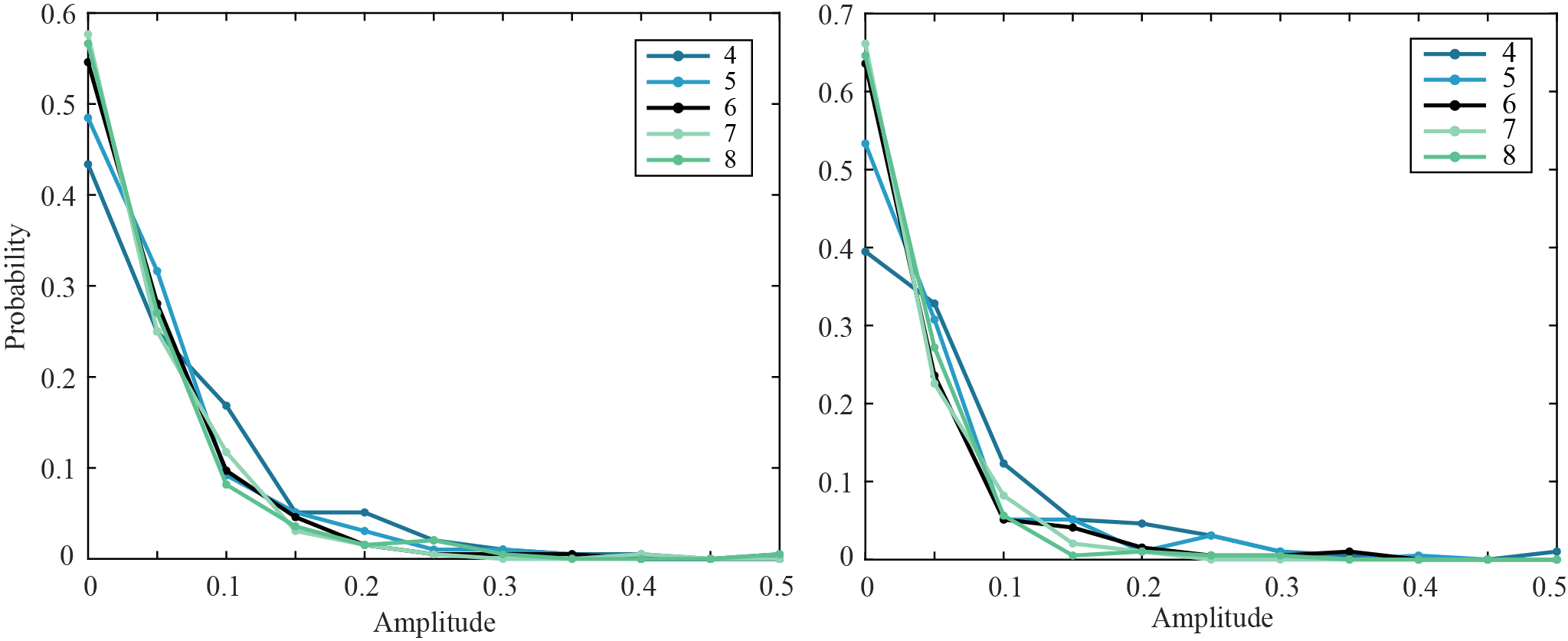}
	\caption{{\bf Probability distributions of amplitude at different output sites for 11-sited (left) and 12-sited (right) lattices.} With the system size increasing, same as the even-sited systems, the odd-sited ring lattice also have a Gaussian amplitude distribution, which means that the Anderson localization exists and helps recover the chiral symmetry in odd-sited systems. The site 4 is the excited site, and others are the neighboring sites.}
	\label{s_11_12_distribution_each_site}
\end{figure*}

The changes in the field amplitude and phase distributions discussed in Ref[S2] can be a good indicator for the Anderson localization shown in main text. To confirm the role of Anderson localization and fully exhibit our experimental results, we show the probability distributions of amplitude for sites, as is shown in Fig.\ref{s_5_6_distribution_each_site} taking 5-/6-sited ring lattices as example. The odd-sited ring lattices have a Rayleigh-like amplitude distribution for sites in the lattice due to the non-chiral symmetry, while the even-sited lattices exhibit Gaussian amplitude distributions. With the system size increasing, as is shown in Fig.\ref{s_11_12_distribution_each_site} for 11-/12-sited ring lattices, same as the even-sited systems, the odd-sited ring lattice also have a Gaussian amplitude distribution, which means that the Anderson localization exists and helps recover the chiral symmetry in odd-sited systems. The result is consistent with the discussion in Ref[S2].

\subsection*{\large VIII. Discussion on experimental imperfection and statistical fluctuation}
As we know, all experimental implementations unavoidably have some imperfections, for example, there are no perfect quantum state preparations and operations. In our experiment, we try our best to lock or optimize all the parameters, including laser pulse energy, repetition rate, beam shaping, angle pointing, writing speed, and even lock the parameters of the fabrication environment, including table vibration, temperature and humidity. There may still exist some imperfections, which may all contribute a simultaneously slight drop of $g^{(2)}$ for all the ring lattices. However, as we can see all the experimental results presented in the main text, the predicted parity-induced gap of $g^{(2)}$ can be well observed.

\begin{figure*}[ht!]
	\centering
	\includegraphics[width=1.2\columnwidth]{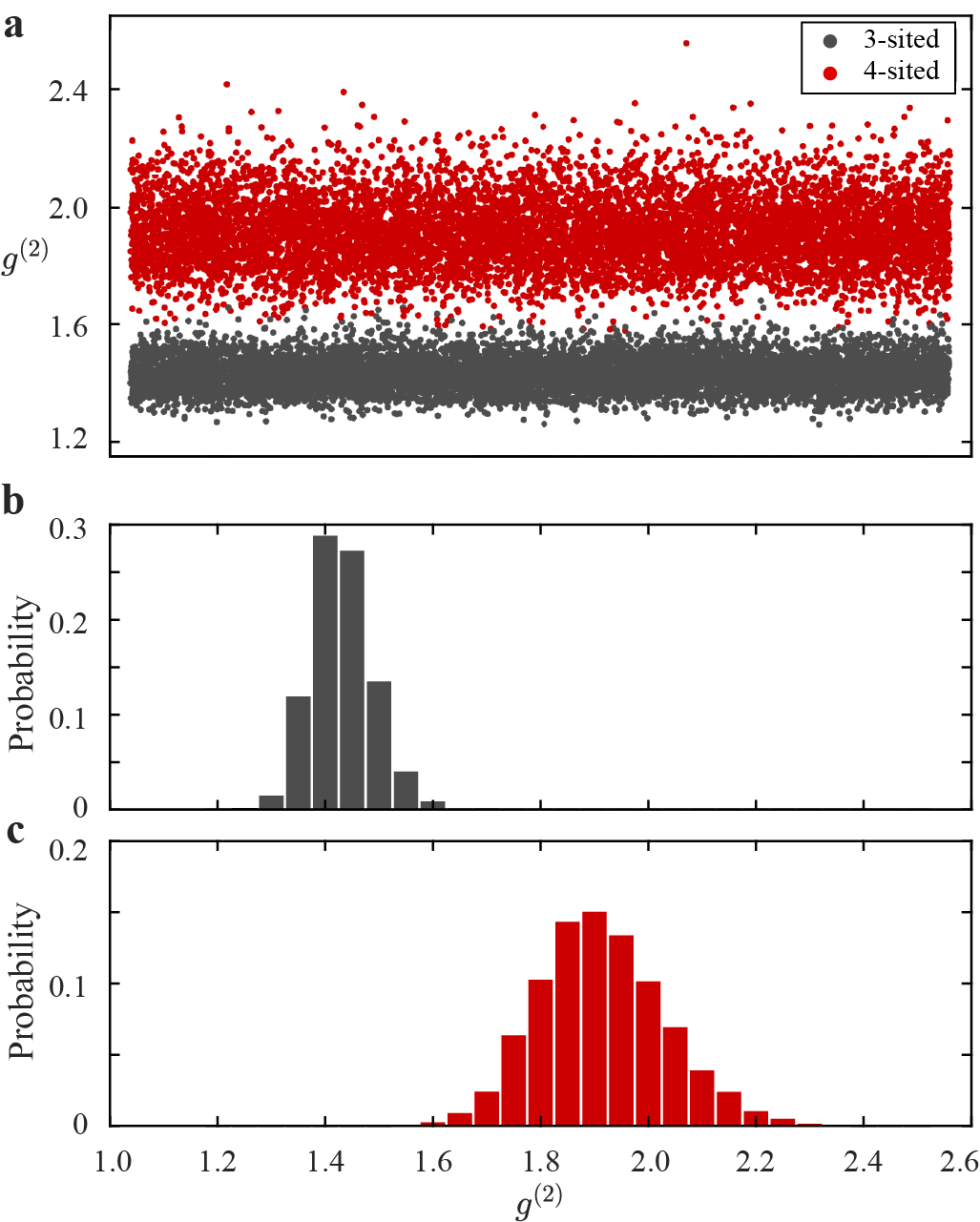}
	\caption{{\bf Specific $g^{(2)}$ value of each simulation and the distribution for 3- and 4-sited ring lattices.} Some points go out of one standard deviation region and therefore the associated difference of $g^{(2)}$ can be smaller or larger than that of averaged $g^{(2)}$. The difference of $g^{(2)}$ can be observed for odd- and even-sited lattices with 120 samples with statistic fluctuation.}
	\label{s_fluctuation}
\end{figure*}

Besides unavoidable imperfection, as is shown in Fig. 2b and 2c of the main text, the difference of $g^{(2)}$ highly depends on the number of samples, where we obtain the result by doing 800 simulated evolution for the same ensemble size. The gray region is one standard deviation away of the statistical average value, which does not show all the simulated results but give a clear instruction for the appropriate number of samples in order to observe the gap of $g^{(2)}$. We visualize all our simulation results of 3- and 4-sited lattices in Fig.\ref{s_fluctuation} showing the specific $g^{(2)}$ value of each simulation and the distribution. We can see that some points go out of one standard deviation region and therefore the associated difference of $g^{(2)}$ can be smaller or larger than that of averaged $g^{(2)}$. The difference of $g^{(2)}$ can be observed for odd- and even-sited lattices with 120 samples, but their values obtained in a certain simulation or experiment not necessarily exactly equal to the difference of averaged $g^{(2)}$.

In summary, our main goal is to demonstrate that the gap of $g^{(2)}$ in even-parity disordered ring lattices is observed for limited system in experiment while strong disorder-induced Anderson localization overwhelm such a phenomenon in large-scale structures. In our experiment, even with unavoidable experimental imperfection and statistical fluctuation, we still are able to observe the predicted parity-induced gap of $g^{(2)}$ for excited sites. 

\begin{figure*}[ht!]
	\centering
	\includegraphics[width=1.8\columnwidth]{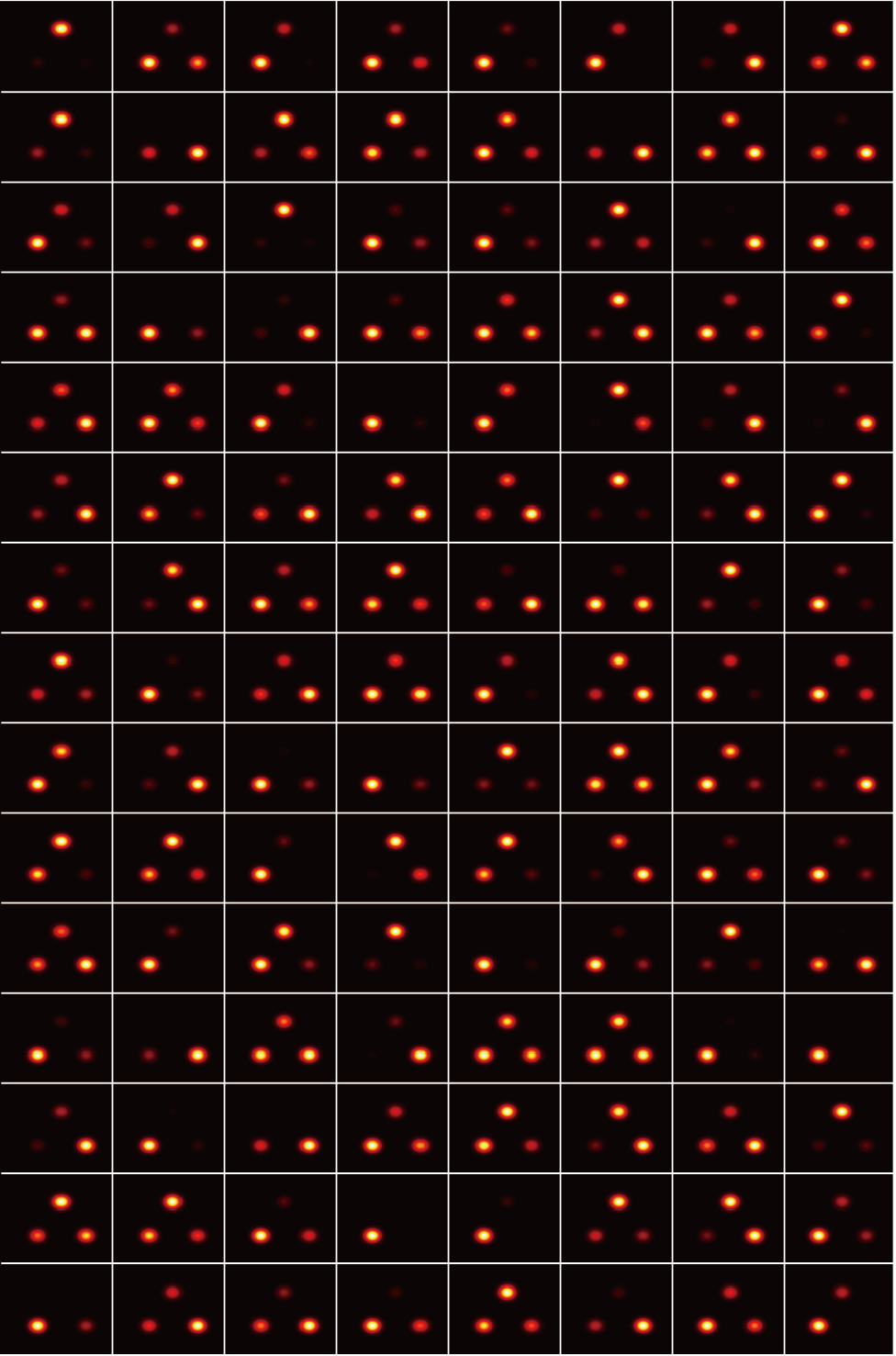}
	\caption{{\bf Experimental amplitude distributions of 120 3-sited ring lattices.}}
	\label{l3}
\end{figure*}

\begin{figure*}[ht!]
	\centering
	\includegraphics[width=1.8\columnwidth]{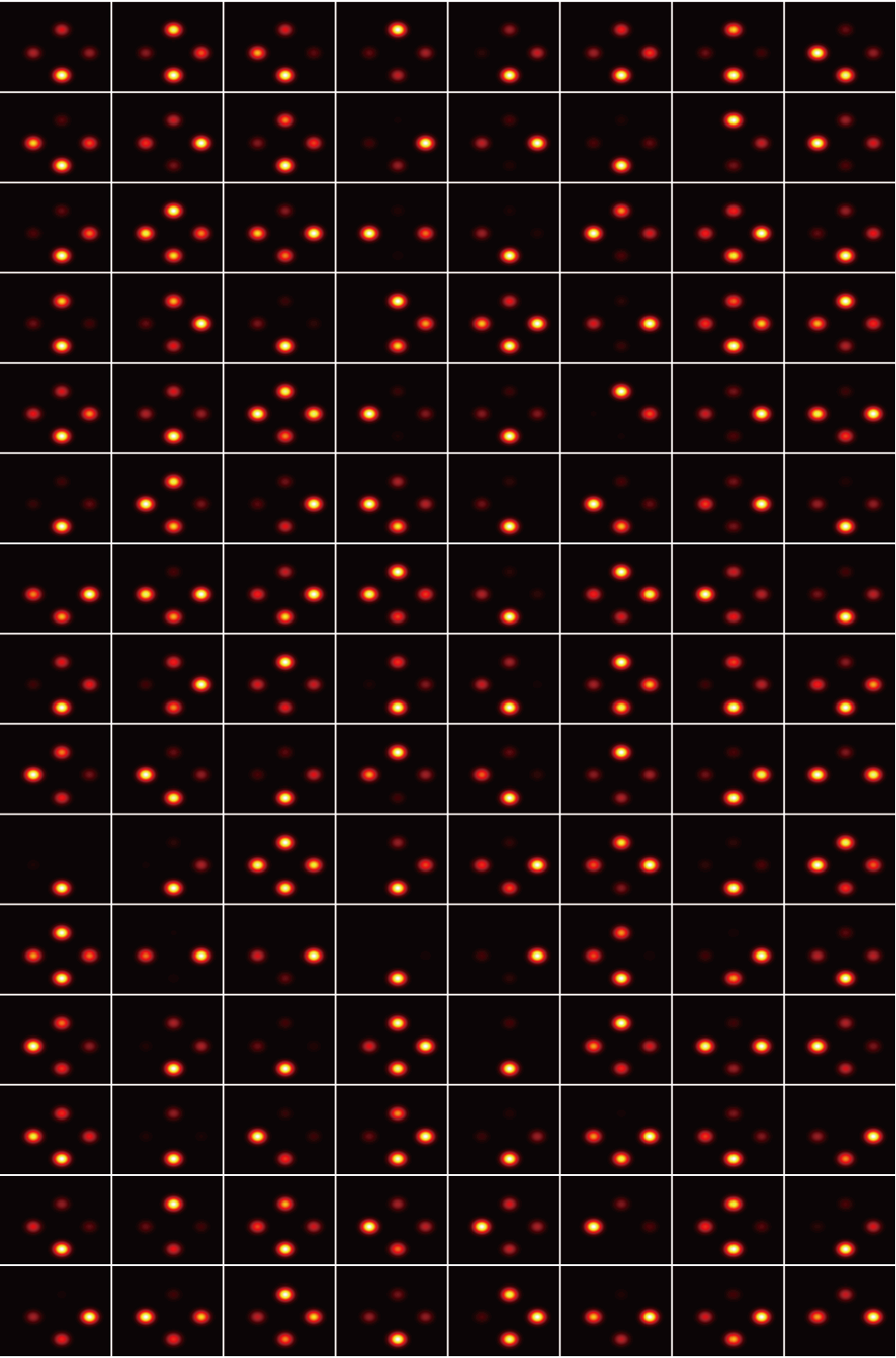}
	\caption{{\bf Experimental amplitude distributions of 120 4-sited ring lattices.}}
	\label{l4}
\end{figure*}

\begin{figure*}[ht!]
	\centering
	\includegraphics[width=1.8\columnwidth]{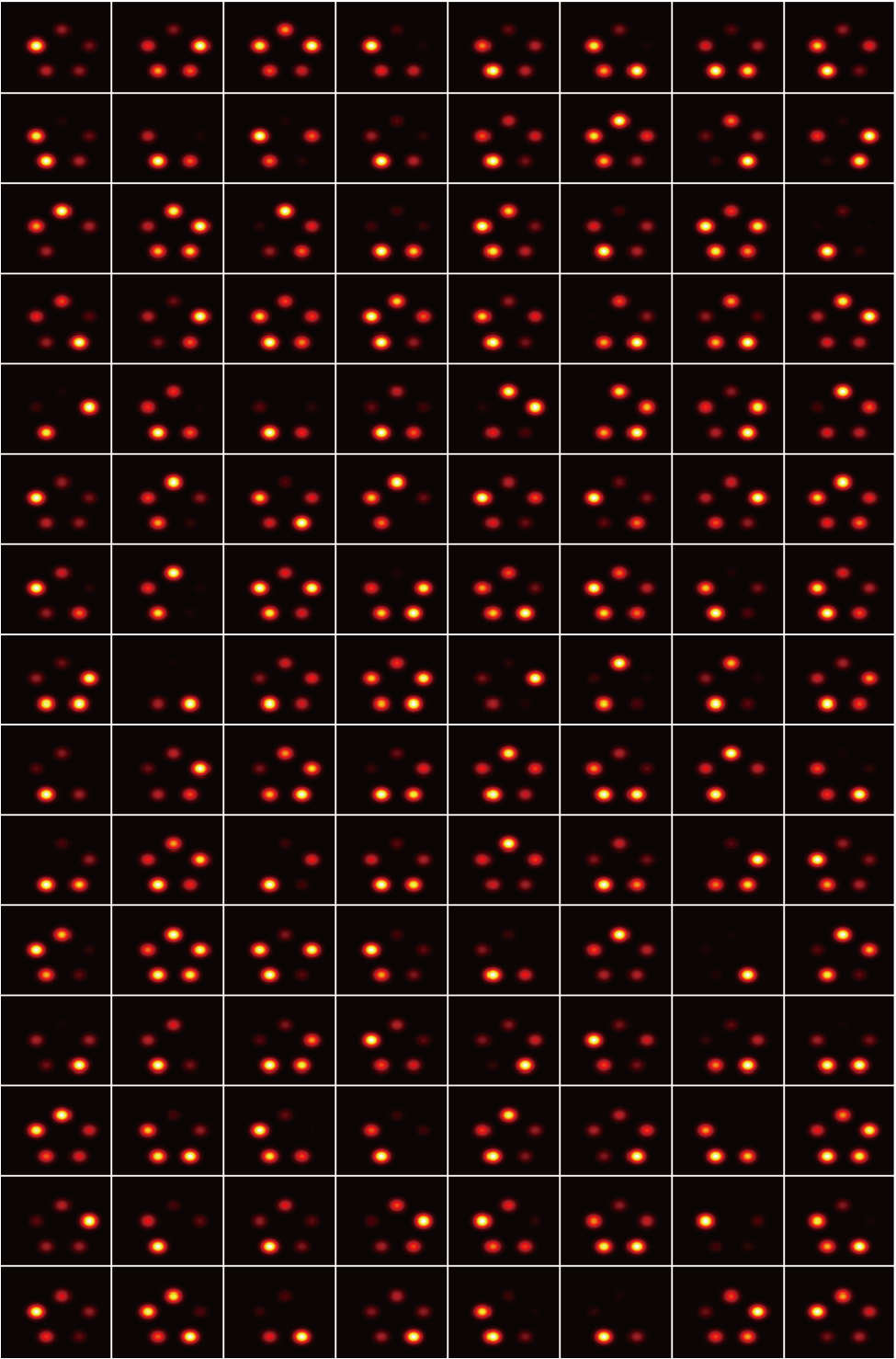}
	\caption{{\bf Experimental amplitude distributions of 120 5-sited ring lattices.}}
	\label{l5}
\end{figure*}

\begin{figure*}[ht!]
	\centering
	\includegraphics[width=1.8\columnwidth]{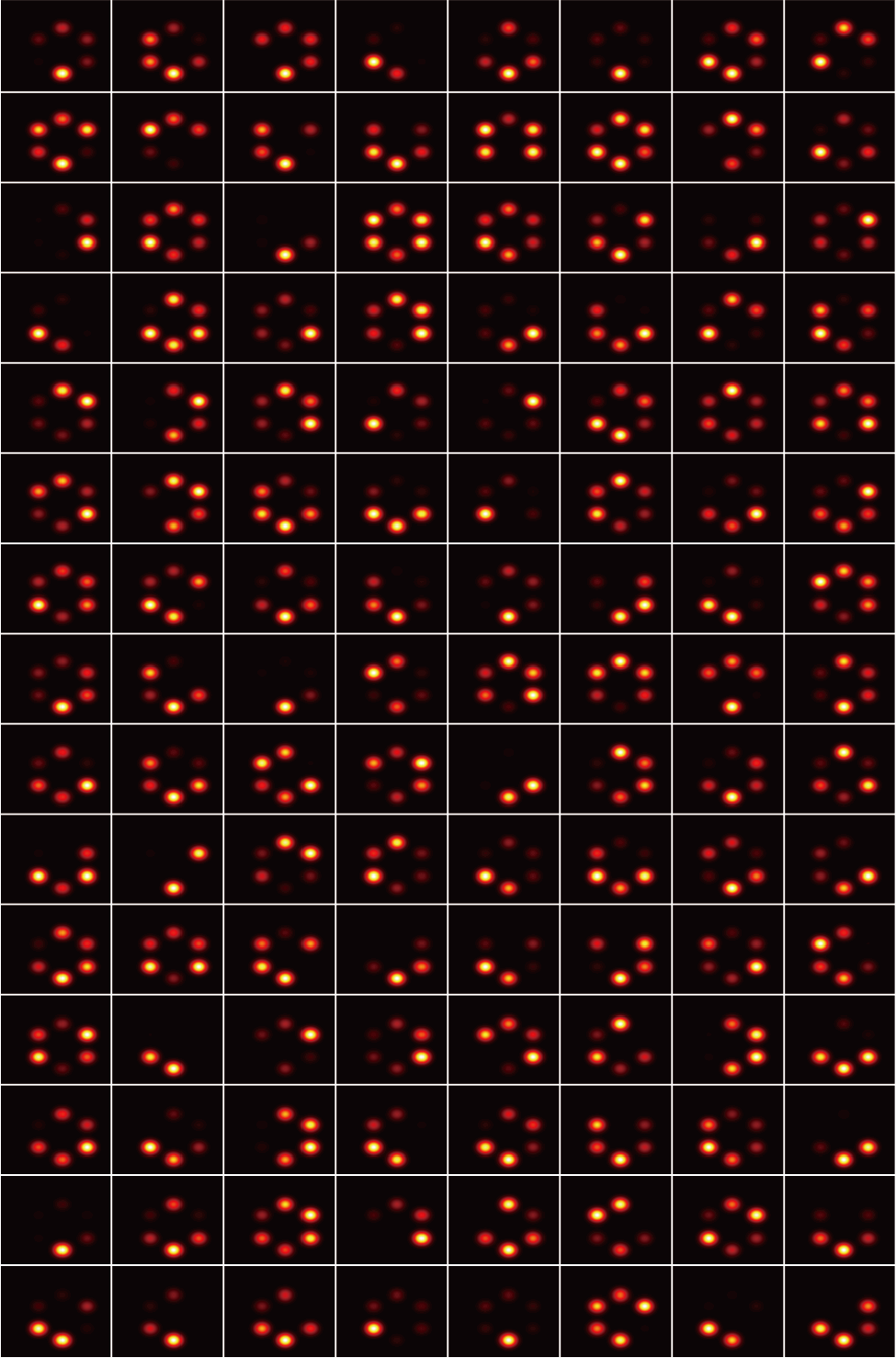}
	\caption{{\bf Experimental amplitude distributions of 120 6-sited ring lattices.}}
	\label{l6}	
\end{figure*}

\begin{figure*}[ht!]
	\centering
	\includegraphics[width=1.8\columnwidth]{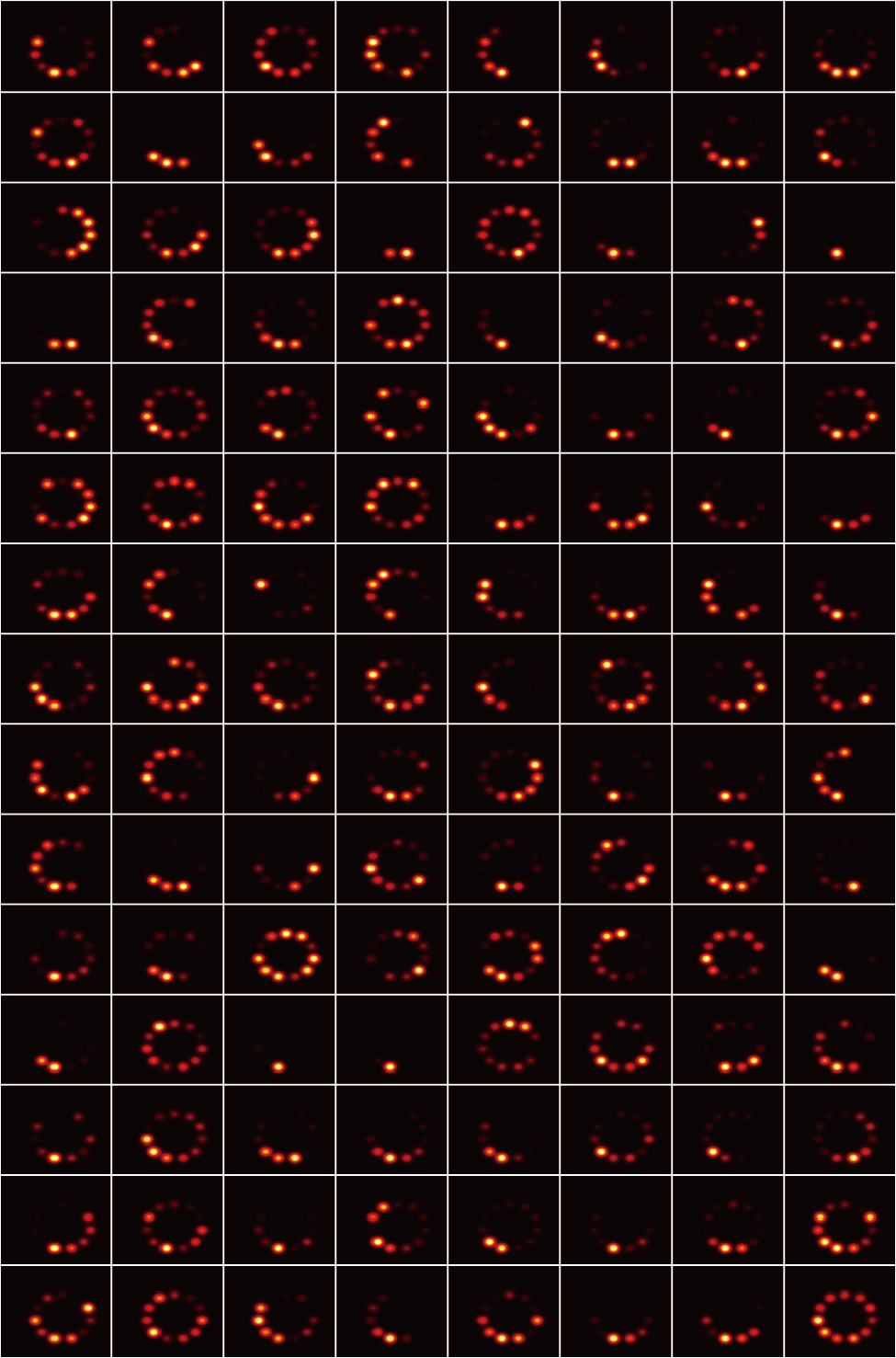}
	\caption{{\bf Experimental amplitude distributions of 200 11-sited ring lattices (Part 1).}}
	\label{11-1}
\end{figure*}

\begin{figure*}[ht!]
	\centering
	\includegraphics[width=1.8\columnwidth]{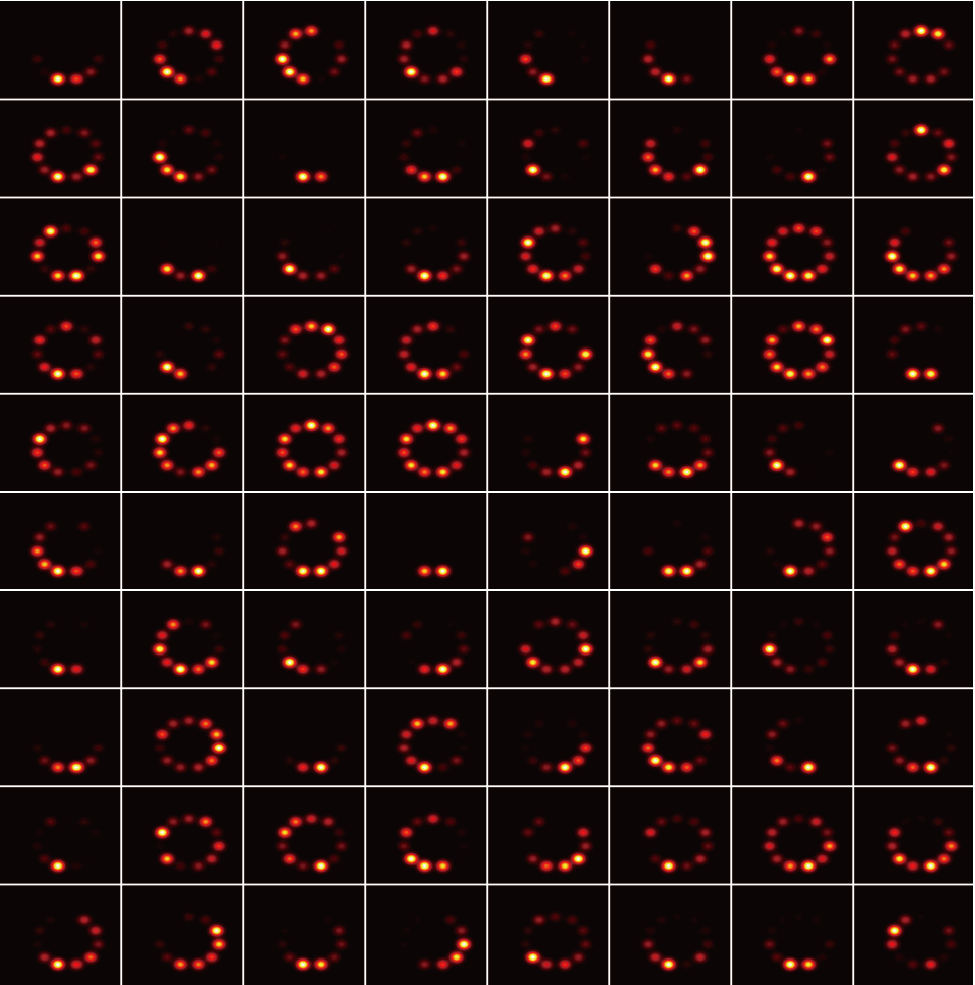}
	\caption{{\bf Experimental amplitude distributions of 200 11-sited ring lattices (Part 2).}}
	\label{11-2}
\end{figure*}

\begin{figure*}[ht!]
	\centering
	\includegraphics[width=1.8\columnwidth]{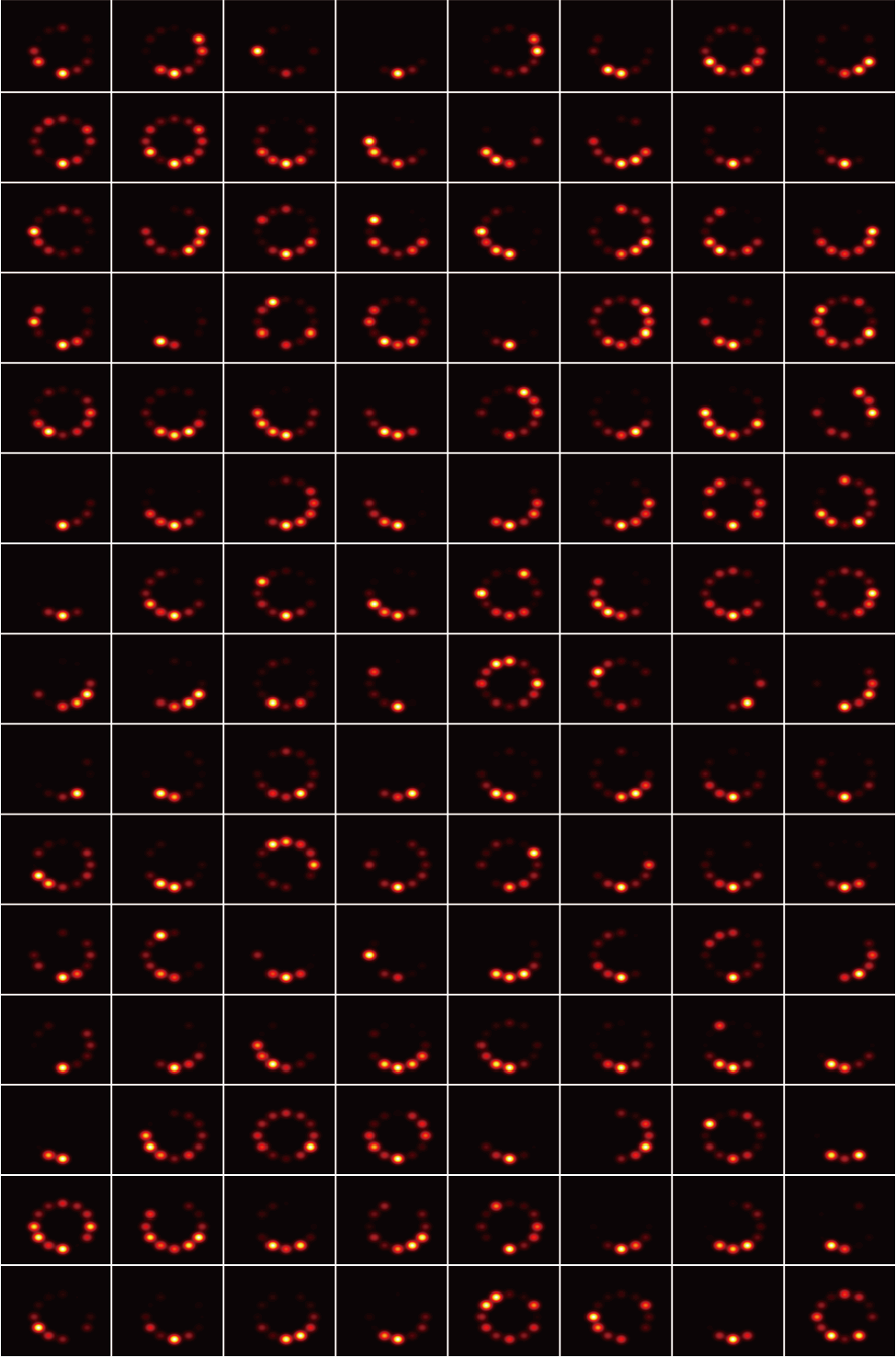}
	\caption{{\bf Experimental amplitude distributions of 200 12-sited ring lattices (Part 1).}}
	\label{12-1}
\end{figure*}

\begin{figure*}[ht!]
	\centering
	\includegraphics[width=1.8\columnwidth]{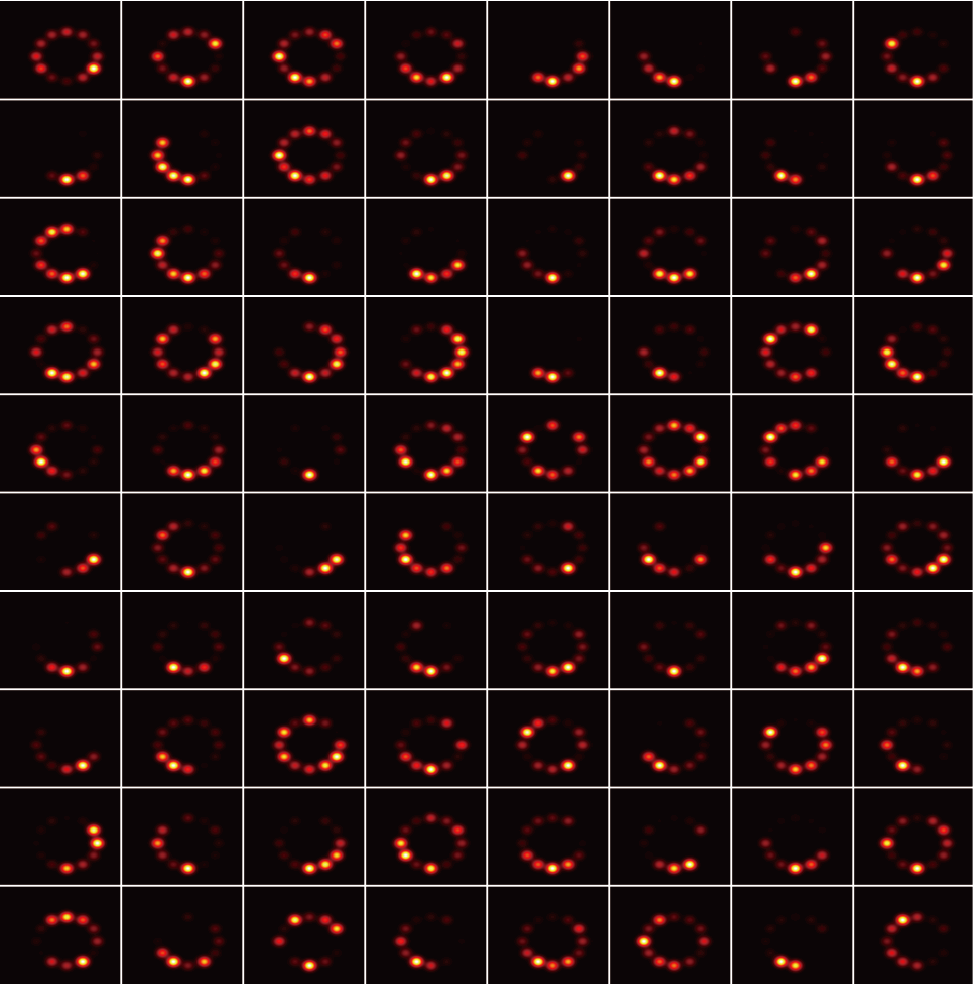}
	\caption{{\bf Experimental amplitude distributions of 200 12-sited ring lattices (Part 2).}}
	\label{12-2}
\end{figure*}

We further show the measured amplitude distributions of all samples in Fig.\ref{l3}--\ref{12-2}, which may also facilitate the audiences for potential investigations in different way.\\

\noindent ------------------------------

\noindent [S1] T. Schwartz, G. Bartal, S. Fishman, and M. Segev, Nature (London) \textbf{446}, 52 (2007).\\
\noindent [S2] H. E. Kondakci, A. F. Abouraddy, and B. E. Saleh, Sci. Rep. \textbf{7}, 8948 (2017).



\end{document}